# CONSTELLATION SHARED MULTIPLE ACCESS - A NOMA SCHEME FOR INCREASED USER CAPACITY IN 5G MMTC

Kiran V. Shanbhag[1] and Savitha H. M.[2]

[1]Department of Electronics and Communication Engineering,
Anjuman Institute of Technology and Management, Bhatkal, Karnataka, India
[2]Department of Electronics and Communication Engineering,
St Joseph Engineering College, Mangaluru, India

## ABSTRACT

*While the legacy cyclic prefix orthogonal frequency division multiple access is retained as the preferred multiple access scheme for 5G enhanced mobile broadband the research is now focussed on the multiple access schemes for massive machine type communication (mMTC) and ultra-reliable low latency communication .Though orthogonal multiple access schemes provide simple reception, they limit number of simultaneous user equipment as against the primary requirement of mMTC. On the other hand, the various non-orthogonal multiple access schemes which have been proposed so far as the likely solution, need complex successive interference cancellation receivers. So a simplified scheme named constellation shared multiple access is proposed here which substantially increases the number of simultaneous users to be served within a single resource block (RB) in LTE or 5G New Radio, thus aiding the massive connectivity requirement of mMTC. This is achieved by differentiating among the users in constellation domain. Moreover, the simple architecture compatible with 5G eMBB makes it a strong contender multiple access contender for 5G mMTC.*

## 1. INTRODUCTION

Fifth generation (5G) wireless network is now almost a commercial reality as the standardization is under process. 5G networks support three major applications, which include enhanced mobile broadband (eMBB), massive machine- type communications (mMTC) and ultra-reliable and low-latency communications (URLLC). These scenarios require massive connectivity, high system throughput and improved spectral efficiency (SE) as compared to long term evolution (LTE) [1]. In order to meet these new requirements, new modulation and multiple access (MA) schemes are being explored. Though CP OFDMA has been almost adapted as the preferred technique for eMBB due to its ability to enhance communication speed, the synchronized nature of the scheme limits the number of User Equipment (UE) that can be simultaneously supported. But one of the key requirements of mMTC and URLLC is the need to support a large number of UEs that not only try to communicate simultaneously but with minimum involved overhead which are the issues that OFDMA cannot address as is. As a result several Non-Orthogonal Multiple Access (NOMA) schemes are being proposed as a solution. The unsynchronized manner in which the schemes work, enables them to not only serve non uniform number of UEs simultaneously but also to provide grantless access which reduces overheads thus improving the latency [2],[3].

 

In a typical NOMA system, signal transmitter and receiver are jointly optimized, so that multiple layers of data from more than one UE can be simultaneously delivered using the same time, frequency and spatial resource in a superimposed manner. At the receiver side, the information of different UEs can be recovered by advanced receivers such as successive interference cancellation (SIC) or iterative detection receivers [4]. This differs from conventional power allocation strategies, such as water filling, as NOMA allocates less power for the users with better downlink channel state information (CSI), to guarantee overall fairness and to utilize diversity in the time /frequency/ code domains. The user with more transmit power is first to be decoded while treating the other user's signal as noise. Once the signal corresponding to the user with the larger transmit power is detected and decoded, its signal component will be subtracted from the received signal by SIC receiver to facilitate the detection of subsequent users. The difference among these schemes is mainly on UE's signature design, i.e., whether the scrambling sequence, interleaving or spreading code is used to differentiate UEs.

### 1.1. Related Work

One of the earliest NOMA schemes proposed is a scheme called sparse code multiple access (SCMA), a CDMA based scheme but with sparse codes and multidimensional constellation to separate simultaneous users in same resource block [5]. But the scheme is prone to user collision needing complex reception under overloaded scenario. Moreover the scheme, like any other multicarrier modulation (MCM) schemes, is suffering from high PAPR and there is no mention of single carrier version as solution. Another prominent scheme presented is named resource spread multiple access (RSMA) which uses low rate channel codes and scrambling with good correlation properties to separate users , also needing SIC receivers [6]. Interleave Division Multiple Access another scheme which uses the concept of interleaving to increase user capacity [7]. Though all these schemes tend to enhance simultaneous user capacity, it should be noted that the superimposed nature of data forces complex iterative reception which may not be efficient in terms of power consumption, receiver hardware complexity as against the requirements of mMTC. A detailed survey of several such NOMA contenders classify them as either scrambling based, spreading based or interleaving based but with SIC based receiver as a common challenge [8]. It's also observed that as more and more users are accommodated, either the performance deteriorates drastically at receiver side or needs complex iterative reception. The recent study has [9] also revealed that 5G systems are likely to operate with higher order modulation schemes such as 1024 QAM, 2048 QAM and 4096 QAM in millimetre wave frequencies without much bandwidth constraint but with high connection density requirements of about 1M devices/Km2. The coexistence of eMBB with mMTC and uRLLC is also a research concern with compatibility being the preference. The 5g services in general are expected provide a wide range of quality of service (QoS) ie provide different priority among different applications in terms of bitrate, latency, jitter etc or to guarantee a certain performance in terms of data rate/ throughput . This is usually done by having a suitable QoS flow identifier (QFI)[10], [11].

In order to address these issues, a new multiple access scheme named constellation shared multiple access (CSMA) is presented here. It uses non overlapping constellation points allocation to differentiate among several simultaneous users within a set of subcarriers in a LTE resource block. This not only enhances the number of users, but also the non-overlapping nature of the resource allocation i.e. constellation points in this case, greatly simplifies the reception process as against SIC reception [12][13]. The scheme can also satisfy wide QoS requirements that arise in mMTC scenarios. The CSMA scheme can provide for the bitrate/ throughput aspects of QoS flow by supporting different guaranteed bit rate (GBR) and non GBR flows as it can flexibly schedule the constellation resources to achieve minimum requirements to certain users.The scheme proposed is a simple extension of legacy CP OFDMA in LTE such that, it allows the existing set of subcarriers in a single resource block(RB) that would have been otherwise used by a single UE in OFDMA block, to be now utilized by multiple UE by separating them in the constellation domain as compared scrambling, interleaving or code domain as suggested in other NOMA approaches. This results in a substantial increase in the number of UEs that can be simultaneously serviced. The scheme also has advantage over other NOMA schemes due to the following reasons. Firstly, it's the similarity and compatibility with OFDMA, which aides in simple migration to NOMA mode when more UEs are present. Secondly, the non-overlapping nature of constellation mapping, which ensures simplicity of the decoding scheme, as compared to the SIC reception in case of other NOMA schemes. Throughout this study, we will assume an underlying OFDMA layer that takes care of synchronization, inter-symbol interference (ISI) etc and the focus will be mainly on accommodating/differentiating users in a single RB built on this OFDMA layer. Section 2 illustrates the proposed scheme, discusses in detail the improvement in user capacity, the various advantages and disadvantages of the scheme. Section 3 includes the simulation results and section 4 includes the conclusion and scope for further research on the scheme.

## 2. PROPOSED SCHEME

The proposed scheme is developed as an extension of existing CP OFDMA based LTE architecture. For orthogonal multiple access like OFDMA, as per the standard terminology, a RB is the smallest unit of resource that can be allocated to a single user[14]. For an FDD (frequency domain duplex) based frame structure, it comprises of 12 subcarriers in frequency domain along with one time slot comprising of either 7 OFDM symbols in LTE or 14 OFDM symbols in 5G NR[15]. While the data from a single UE is usually QAM modulated to be mapped onto an entire subcarrier resource block, the idea in proposed scheme is to accommodate multiple users that are separated in QAM constellation, in the same RB in a time shared manner. This is done by mapping the data bits of individual UE to a higher constellation by using either a look up table or suitable mapping logic.

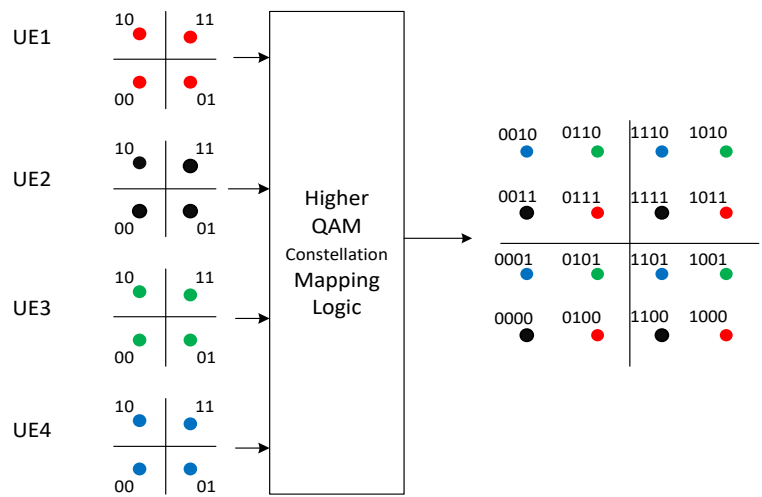

Figure 1. A sample lookup table based higher constellation mapping scheme where four UE data with 4 QAM constellation is being mapped to a single 16 QAM constellation

An example scheme is depicted in figure 1, where each 2 bit symbol data from 4 users, which would need a QAM modulation with 4 constellation points, have been mapped to 4 bits such that the resulting 16 constellation points are equally divided among 4 users, as per the look up table 1. This should allow a single RB with 16 QAM modulated OFDM symbols, to serve four.

Table 1. Look up table for the scheme in figure 1, where each 2 bit symbol data of 4 UEs are being mapped to unique 4 bit symbol data of 16 QAM constellation

| User | User Data | Corresponding Higher Constellation Point |
|---|---|---|
| UE1 | 00 | 0100 |
|     | 01 | 1000 |
|     | 10 | 0111 |
|     | 11 | 1011 |
| UE2 | 00 | 0000 |
|     | 01 | 1100 |
|     | 10 | 0011 |
|     | 11 | 1111 |
| UE3 | 00 | 0101 |
|     | 01 | 1001 |
|     | 10 | 0110 |
|     | 11 | 1010 |
| UE4 | 00 | 0001 |
|     | 01 | 1101 |
|     | 10 | 0010 |
|     | 11 | 1110 |

UE, that are separated in constellation space. To avoid additional signalling overhead per user, it is assumed that all the users sharing the RB experience similar channel conditions. Another approach for mapping onto the higher constellation would be to simply combine unique user identifier (address) bits along with their data bits in a particular bit location so as to share QAM constellation points equally among several UEs. Say, if the total available constellation points at any instant in M-QAM such that $M = 2^D$, then these are to be shared among several UEs in a systematic manner such that fixed 'A' bits out of 'D' are used for addressing 2A user equipment and remaining bits are used for actual data. If, for example, 64 QAM is being used, and the resources are to be shared among 4 users, then each UE gets 16 unique constellation points, amounting to 4 data bits per user. Then out of six bits which would form $2^6$ i.e. 64 constellation point, any 2 bits are used as part of address of UE and remaining 4 bits are allocated for actual data. Similarly, if any 3 bits out of 6 bits are used to identify UEs, the scheme can increase simultaneous user capacity 8 times and so on. The table 2 depicts one of the schemes for 64 QAM where, out of 6 bits, the central 2 bits represent one of the 4 UE's addresses, whereas first and last 2 bits represent the actual data of corresponding user. Figure 2 shows the corresponding constellation diagram where the UEs are sharing all 64 constellation.

Table 2. An example scheme where out of 6 bits constituting a constellation point, 2 bits in centre indicate UE address and remaining bits indicate the actual UE data

| $D_5$ $D_4$ $D_3$ $D_2$ $D_1$ $D_0$ | $D_5$ $D_4$ $D_3$ $D_2$ $D_1$ $D_0$ | $D_5$ $D_4$ $D_3$ $D_2$ $D_1$ $D_0$ | $D_5$ $D_4$ $D_3$ $D_2$ $D_1$ $D_0$ |
|---|---|---|---|
| $B_3$ $B_2$ $A_1$ $A_0$ $B_1$ $B_0$ | $B_3$ $B_2$ $A_1$ $A_0$ $B_1$ $B_0$ | $B_3$ $B_2$ $A_1$ $A_0$ $B_1$ $B_0$ | $B_3$ $B_2$ $A_1$ $A_0$ $B_1$ $B_0$ |
| 0 0 0 0 0 0 | 0 0 0 1 0 0 | 0 0 1 0 0 0 | 0 0 1 1 0 0 |
| 0 0 0 0 0 1 | 0 0 0 1 0 1 | 0 0 1 0 0 1 | 0 0 1 1 0 1 |
| x x 0 0 x x | x x 0 1 x x | x x 1 0 x x | x x 1 1 x x |
| x x 0 0 x x | x x 0 1 x x | x x 1 0 x x | x x 1 1 x x |
| 1 1 0 0 1 0 | 1 1 0 1 1 0 | 1 1 1 0 1 0 | 1 1 1 1 1 0 |
| 1 1 0 0 1 1 | 1 1 0 1 1 1 | 1 1 1 0 1 1 | 1 1 1 1 1 1 |
| **UE 1** data combined with address bits **0 0** | **UE 2** data combined with address bits **0 1** | **UE 3** data combined with address bits **1 0** | **UE 4** data combined with address bits **1 1** |

points equally among each other. Thus the 4 UEs can now share the same RB in a time shared manner, though with reduced throughput. As indicated by the diagram, the constellation points corresponding to individual UE are clearly differentiated from each other and do not overlap. Both the mapping schemes, the one with the look up table and the illustration with 2 centre bits used as user address are straight forward simple schemes among several possibilities, meant for the readers understanding, but may not be an ideal one. The optimum allocation and the distribution of user constellation points should be done by taking into consideration, the shaping gain as well as PAPR experienced by each UE.[16], [17]

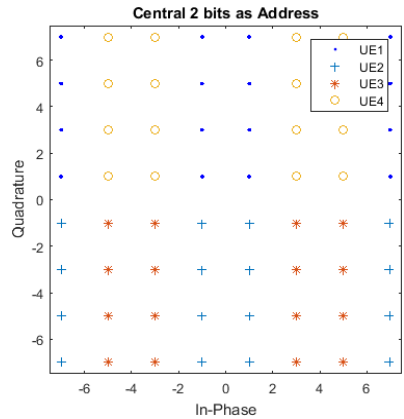

Figure 2. The CSMA scheme as per table 2, where 64 QAM constellation points are shared by 4 UEs as per legend

## 2.1. Features and limitations

The following section describes the advantageous features of the scheme such as user capacity expansion, provision of QoS and also the impact of the scheme on the throughput.

### 2.1.1. User Capacity Expansion

Depending on the existing channel conditions and available SNR, which in turn determine the kind of modulation scheme to be used, the numbers of concurrent users supported also vary. Table 3 summarizes likely increase in capacity (Uc) for a particular modulation scheme, based on 'B', the symbol width of actual data in individual UEs and 'M' the higher constellation scheme employed for multiplexed data , given by the expression

$$U_c = 2^{(\log_2 M - B)} \quad (1)$$

Table 3. The User capacity enhancement factor for various modulation schemes and different symbol widths of the UE

| Modulation Scheme M | Symbol width of UE 'B' bits | User capacity Enhancement Factor, Uc |
|---|---|---|
| 64 QAM | 4 | 4 |
| 64 QAM | 3 | 8 |
| 64 QAM | 2 | 16 |
| 256 QAM | 6 | 4 |
| 256 QAM | 4 | 16 |
| 256 QAM | 2 | 64 |
| 1024 QAM | 8 | 4 |
| 1024 QAM | 6 | 16 |
| 1024 QAM | 4 | 64 |

### 2.1.2. Provision of Quality of Service (QoS)

The scheme can also be used to provide different QoS among shared UEs by doing non uniform constellation scheduling similar to guaranteed bit rate (GBR) bearer. There are two means of doing so. One is by having a look up table with more points allocated to particular UE along with allocating relatively more OFDM symbols per slot per RB for such UE with higher throughput requirements. Figure 3 shows one such scheme where where out of 16 constellation points, UE1 has been allocated 8 , thus being able to send 3 information bits per symbol while UE2, UE3 get 4 points each and are able to send only 2 bits . Table 4 shows the corresponding look up table. Secondly, the bit allocation for address and data can be changed dynamically as per the QoS requirements for a certain group of UEs, which can be together multiplexed to a common QAM modulator block. This way, while few of 64 QAM modulators can serve say, 4 UE by using 4 bit symbol width for better throughput, remaining can serve 8 UE by using 3 bit symbol width for with reduced throughput.

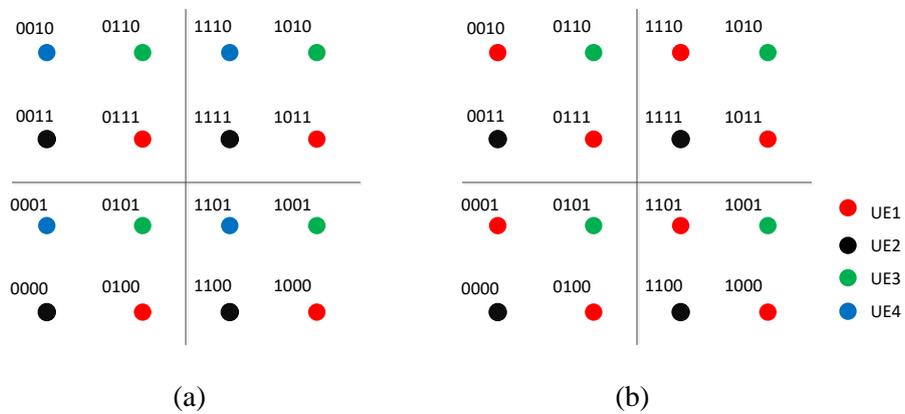

Figure 3. A comparison of uniform constellation sharing among users in (a)  vs non uniform constellation sharing in (b) where UE1 has more priority

Table 4.  Look up table where UE1 has high throughput compared to UE2, UE3

| User | User Data | Corresponding Higher Constellation Point |
|---|---|---|
| UE1 | 000 | 0100 |
|  | 001 | 1000 |
|  | 010 | 0111 |
|  | 011 | 1011 |
|  | 100 | 0001 |
|  | 101 | 1101 |
|  | 110 | 0010 |
|  | 111 | 1110 |
| UE2 | 00 | 0000 |
|  | 01 | 1100 |
|  | 10 | 0011 |
|  | 11 | 1111 |
| UE3 | 00 | 0101 |
|  | 01 | 1001 |
|  | 10 | 0110 |
|  | 11 | 1010 |

### 2.1.3. Reduced receiver complexity

Most of NOMA schemes proposed as of now, rely on successive interference cancellation (SIC) receivers which are fairly complex, iterative in nature and are likely to be less power efficient as they need knowledge of channel state information for differential resource allocation, interconnection among all UE for subsequent reception, thus affecting the concern of long battery life in mMTC devices [18]. The proposed scheme, even though requires entire subcarrier data to be processed by all UE, since they are well separated in constellation space, does not need of SIC reception. The receiver complexity would be that of conventional OFDMA receiver itself, as per our earlier assumption along with an additional reference to the lookup table after QAM demodulation by each UE to receive corresponding data.

### 2.1.4. Compatibility with existing platforms and future technologies

While most of the NOMA schemes suggested so far go for mostly clean slate approach as compared to 5G eMBB, needing additional infrastructure for those applications which need high UE support, the proposed scheme can be easily adapted with slight modifications to legacy OFDMA infrastructure. Depending on the scenario, the switch between limited user, bandwidth extensive applications and the massive connectivity low bit rate applications can be easily achieved using suitable control signals. As depicted in fig 1, fig 3, all of the QAM constellation points, along with a RB can be either allocated to single user in an orthogonal manner giving high bit rate or it can be shared in time/constellation space to accommodate more UE. With the introduction of OFDMA in WiFi 6, which is another platform trying to capture the IoT space [19][20], the suggested scheme can be easily extended to address massive connectivity issues.

### 2.1.5. Impact on throughput and bit error rate (BER)

The increased user capacity comes at the cost of reduced data rate as fewer bits are now available for the actual data due to addressing overhead, compared to CP OFDMA case and each shared user is transmitting in a multiplexed manner. But as the target mMTC applications are more focussed towards massive UE connectivity and not necessarily on high bandwidth applications as compared to eMBB, the reduced bit rate is not of much concern as the likely bit rates in mMTC can go as low as few hundred bits per second[21].

The throughput reduction is mainly due to

1. The addressing overhead of 'A' bits out of $\log_2 M$ available bits for each UE identification reduces bit rate by factor $(\log_2 M - A)/\log_2 M$, which further reduces with increasing users, though not in an exponential manner.
2. The multiplexed nature of user data transmission, which forces only a single user among $2^A$ UE connected to common modulator, to transmit/receive data at any instant of time.

Together for an M QAM scheme, out of 'D' bits, for 'A' number of bits used for addressing, the total reduction in throughput can be given by

$$T_r = (\log_2 M - A)/\left((\log_2 M) * 2^A\right) \tag{2}$$

The BER performance of the scheme is relatively low as compared to its orthogonal counterpart where a single UE is connected to a dedicated M QAM modulator. This is due to the fact that out of 'D' bits forming an M QAM symbol, only 'B' bits carry actual data. The addition of more and more addressing bits to accommodate more UEs, forces the use of higher order modulation, which usually have poorer performance, due to the proximity among the constellation points pertaining to other UE. The probability of symbol error (Ps) or the symbol error rate (SER) for any M QAM modulation scheme is given as under

$$P_s = 1 - \left(1 - \frac{2\sqrt{M}-1}{\sqrt{M}} Q\left[\sqrt{\frac{3\upsilon_s}{M-1}}\right]\right)^2 \qquad (3)$$

Here $\vartheta_s$ is the SNR per symbol given as $\vartheta_s = \frac{\varepsilon_s}{\eta_0}$ and $\varepsilon_s$ is energy per symbol and $\eta_0$ is the noise power. It is also related to SNR per bit given by $\vartheta s = \vartheta b \log_2 M$ where $\vartheta_b = \frac{\varepsilon_b}{\eta_0}$ and $\varepsilon_b$ is energy per bit. Say, if for same throughput from a single user perspective, if B bits are allocated for data, then $M=2^B$ QAM is used and the expression for Ps in terms of symbol width B is given as

$$P_s = 1 - \left(1 - \frac{2\sqrt{2^B}-1}{\sqrt{2^B}} Q\left[\sqrt{\frac{3\upsilon_s}{2^B-1}}\right]\right)^2 \qquad (4)$$

However, if 'A' additional bits are appended to accommodate more users, $2^{(B+A)}$ QAM is used, which will have relatively poor performance due to denser constellation point arrangement given in terms of symbol and address bit width by Ps'

$$P_{s\prime} = 1 - \left(1 - \frac{2\sqrt{2^{B+A}}-1}{\sqrt{2^{B+A}}} Q\left[\sqrt{\frac{3\upsilon_s}{2^{B+A}-1}}\right]\right)^2 \qquad (5)$$

## 2.2. Comparison with other NOMA schemes

Most of existing NOMA schemes can be classified as either code domain NOMA (CD-NOMA) or power domain NOMA (PD-NOMA). The intention of most of these schemes is to provide massive connectivity and low latency via grantless access. As compared to the proposed scheme, the following observations are to be made with respect to existing NOMA schemes.

### 2.2.1. Relative latency and power consumption

The grantless access in most NOMA schemes is owing to the fact that no access grant signalling is needed by each UE. But to be able to decode the data pertaining to the last user, all the previous data pertaining to the earlier users shall be recovered iteratively. This SIC reception will in turn reduce battery life which worsens with increasing users. In the proposed scheme however, once the number of users are determined, a RB is granted in time shared manner, without the need for any SIC reception. This scheme, even though slows down the transmission, ensures that a fixed number of users per RB are served, with uniform resources reserved per users, providing almost uniform BER performance, requiring simple reception/transmission and uniform processing time.

### 2.2.2. Impact of higher UE on performance

Figure 4 shows proposed scheme with 4 users sharing a 16 qam constellation with 4, 4 QAM constellation points each. It clearly shows that per user performance will drop to that of a 16 QAM per user, even though its assigned only 4 constellation points. An important point here is to note that for CSMA constellations pertaining to 4 users sharing single RB, even though as a whole equals that of a 16 QAM, since at any instant within a particular time slots, for a particular user there can only be 4 constellation possibilities as in 4 QAM making it relatively robust. Compare it to figure 5 which shows a mere 3 user PD NOMA constellation resembles

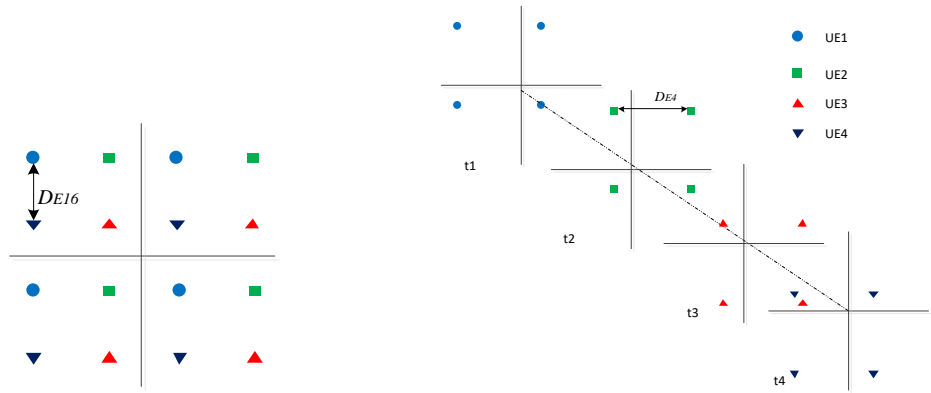

Figure 4. An example 16 QAM constellation being shared by 4 UE, each with 4 points (left) , in time multiplexed manner with t1,t2,t3,t4 representing different intervals within a time slot where they transmit, with each user constellation points enjoying uniform distance from each other (right)

almost a 64 QAM, moreover requiring SIC reception. While most literatures pertaining to NOMA illustrate 2 user cases, higher user numbers reveal another downside. Here, each user data is supposed to be 4 QAM modulated and superimposed in power domain. Those users who do the initial QAM decoding are much reliable as the minimum Euclidian distance, DE4 among corresponding constellation is much higher as shown in figure 5.b. But for the subsequent users, for same bit rate, the constellation points get much closer and closer, affecting the BER performance. As seen from fig 5.c. for UE2, the constellation resembles to that of an inferior 16 QAM with distance DE16. Again from fig 5.d the constellation for UE3 is much inferior resembling that of 64 QAM with distance DE64 thus relying heavily on the error control codes for proper decoding. Having 4th user will make it much worse. This decay with increasing users is much faster in PD- NOMA compared to proposed scheme , as a 4 user scheme CS NOMA is doing much better compared to a mere 3 user PD NOMA, that too without needing SIC.

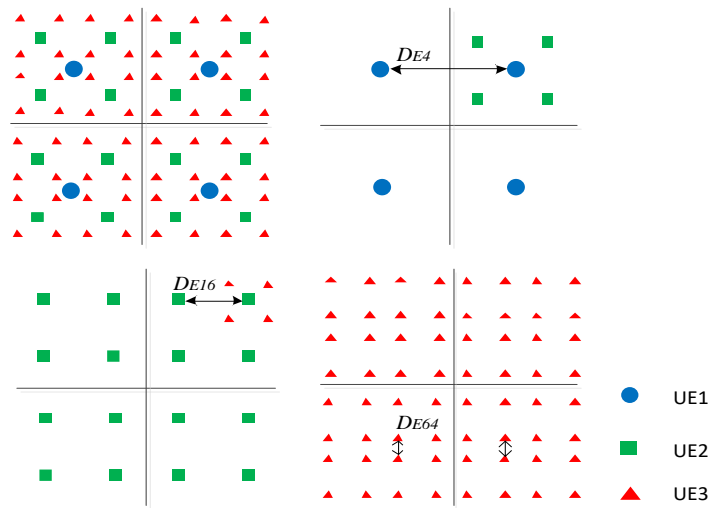

Figure 5. An example 3 user PD NOMA constellation space (top left) with each user employing 4 QAM, with illustrations of user 1 space, along with relatively low power user 2 signal on top (top right), user 2 constellation space, along with low power user 3signal on top of it (bottom left) clearly depicting Euclidean distance reduction with increasing users

### 2.2.3. Ease of scalability

While in most NOMA, accommodating more number of users involve either a robust CSI feedback mechanism in case of PD NOMA [22] or an elaborate codebook design in case of SCMA [23][24] and so on, for the proposed scheme however, the addition of users happens in an exponential manner with relative ease as per a look up table or mechanism discussed in section 2.1.1, with provision to even accommodate various QoS with ease.

## 3. METHOD

The downlink transmitter block diagram of the proposed multiple access schemes is as shown in figure 6. It's a simple extension of the conventional CP OFDMA transmitter architecture for LTE except for the initial multiplexing stage represented by the dashed area. In this initial stage, the actual data bits corresponding to maximum of N simultaneous users such that N=2A are mapped to a higher QAM constellation either by suitable constellation look up table (LUT) mapping or combined with 'A' address bits and fed to one of the K QAM modulators multiplexed in a time shared manner, as against LTE where each modulator would be used for only single user. Thus, the block which would earlier accommodate K users can now accommodate K*N number of users.

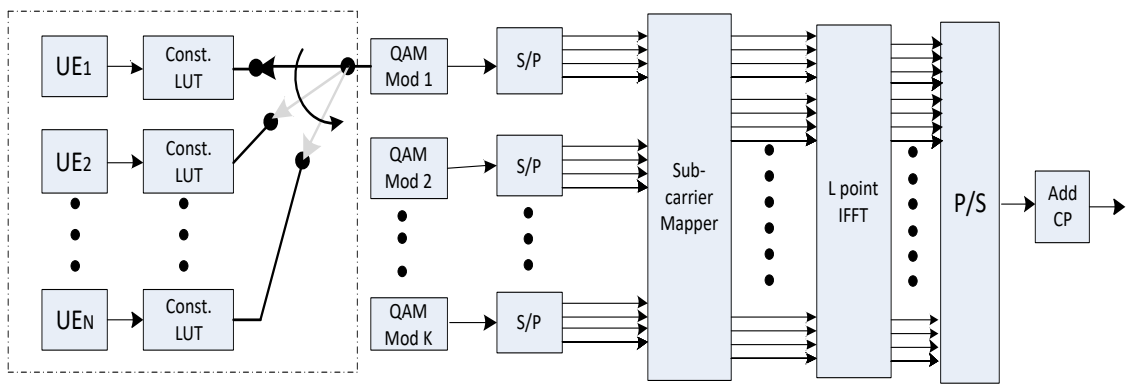

Figure 6. A typical set up of the CSMA scheme for the downlink, which is very similar to CP OFDMA, based LTE Transmitter from the QAM modulator onwards

This higher constellation mapped data is fed to the QAM modulator either based on priority or uniform time shared manner. Here on, the combined user data will be treated as if it were a single UE data and transmitted as OFDMA signal. At the receiver, a reverse process is carried out, where all the data corresponding to a particular demodulator is decoded by all connected users and separated among several UEs once the QAM demodulation is carried out, by identifying the location of the constellation points. Though separated in constellation space, since the same set of subcarriers is shared by several UEs, the scheme is no more an orthogonal scheme, but a NOMA scheme. While 5g eMBB employs single carrier version SCFDMA for uplink transmission to address the peak to average power (PAPR) issue [25],[26], CSMA is also easily adaptable into the single carrier version with little modifications for the uplink transmission. These features make it a strong contender for the 5G, along with CP OFDMA. While usually in LTE or even in 5G NR, a RB is the smallest resource that can be allocated to a single user as in [27],[28],[29] this new approach can accommodate more users per RB by allocating the OFDM symbols within a slot among a group of users, thus increasing the user capacity. Figure 7 shows a sample 5G NR frame structure where a RB in slot 3 containing 14 symbols per slot, is being shared among users indicated by different colours. The users can either alternatively send their OFDM symbols in different time instants within a single slot or in any order as they can be identified by their address bits i.e. constellation location.

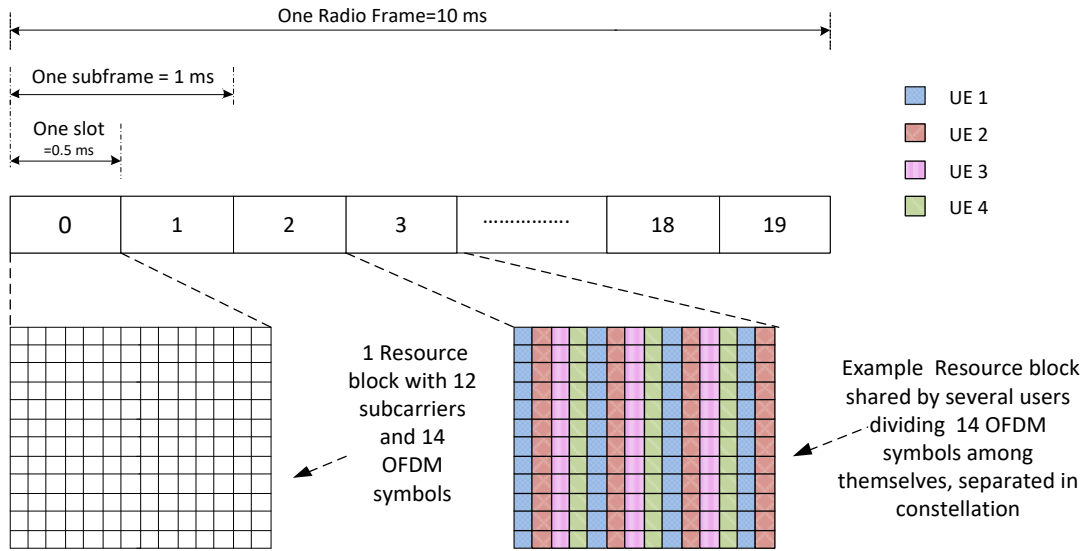

Figure 7. An example 5G frame structure depicting the means of sharing a single RB among several users with each being assigned alternate OFDM symbols within a slot

Number of concurrent users per RB and the possible increase in the number of users is limited by either the order of the highest QAM constellation used based on SNR availability or the number of OFDM symbols per slot in a resource block, if at all the existing 5G NR frame structure architecture is to be used.

## 4. SIMULATION RESULTS

Figure 8 depicts several of the constellation mapping schemes for 64 QAM where out of 6 bits, 2 bits are used for addressing. Matlab software is the tool used for the simulation purpose. While figures 8(a), 8(b) show the constellation sharing obtained by appending 2 address bits to 4 data bits and prepending 2 address bits to data bits respectively, figures 8(c) is obtained by suitable lookup table mapping and 8(d) is obtained by using first and fourth bits as addressing bits. It's observed that while few schemes distribute points uniformly over entire space, few have constellation points in separate quadrants and so on. However, the look up table based approach, allows much more flexibility in constellation allocation. The constellation point mapping among various UE under a single modulator can be devised, taking into consideration several performance parameters such as required quality of service (QoS), Euclidean distance for better differentiation among UE, the spectral efficiency due to phase transitions, constellation shaping gain etc. It's observed that the users can be separated easily and each UE receivers are supposed to receive data corresponding to relevant points after demodulating entire data. A better constellation mapping would be the one which will have all the points distributed with maximum distance, for same user and distributed in entire space so that even if all shared users are not present, the data can be better differentiated from each other with better BER[30].

Figure 9 shows the simulation results of performance comparison of individual UE sharing the RB with three more users with that of a single UE using entire RB as per the CSMA scheme, which clearly shows the faithful recovery of sent data but at the cost of additional SNR requirement. Normalized BER was used instead of SER for fair comparison. 256 point FFT was used for the OFDM with AWGN. No pulse shaping, equalization and error control coding have been employed during simulation. For the conventional OFDMA case with single user per RB , 16 QAM was used and assuming 4 UE sharing same modulator, requiring 2 additional bits along with 4 data bits, 64 QAM was used with 6 bit symbol  Hence, while the former performed better due to lower modulation order, the latter experienced a denser constellation space with reduced Euclidean distance among points as shown in figure 10,  due to the increased modulation order, as evident in the BER curve.  Though there can be several factors to be taken into consideration in choosing suitable mapping, most of them perform almost similar in terms of BER performance as all receivers are required to demodulate under highest QAM order. This is depicted in fig

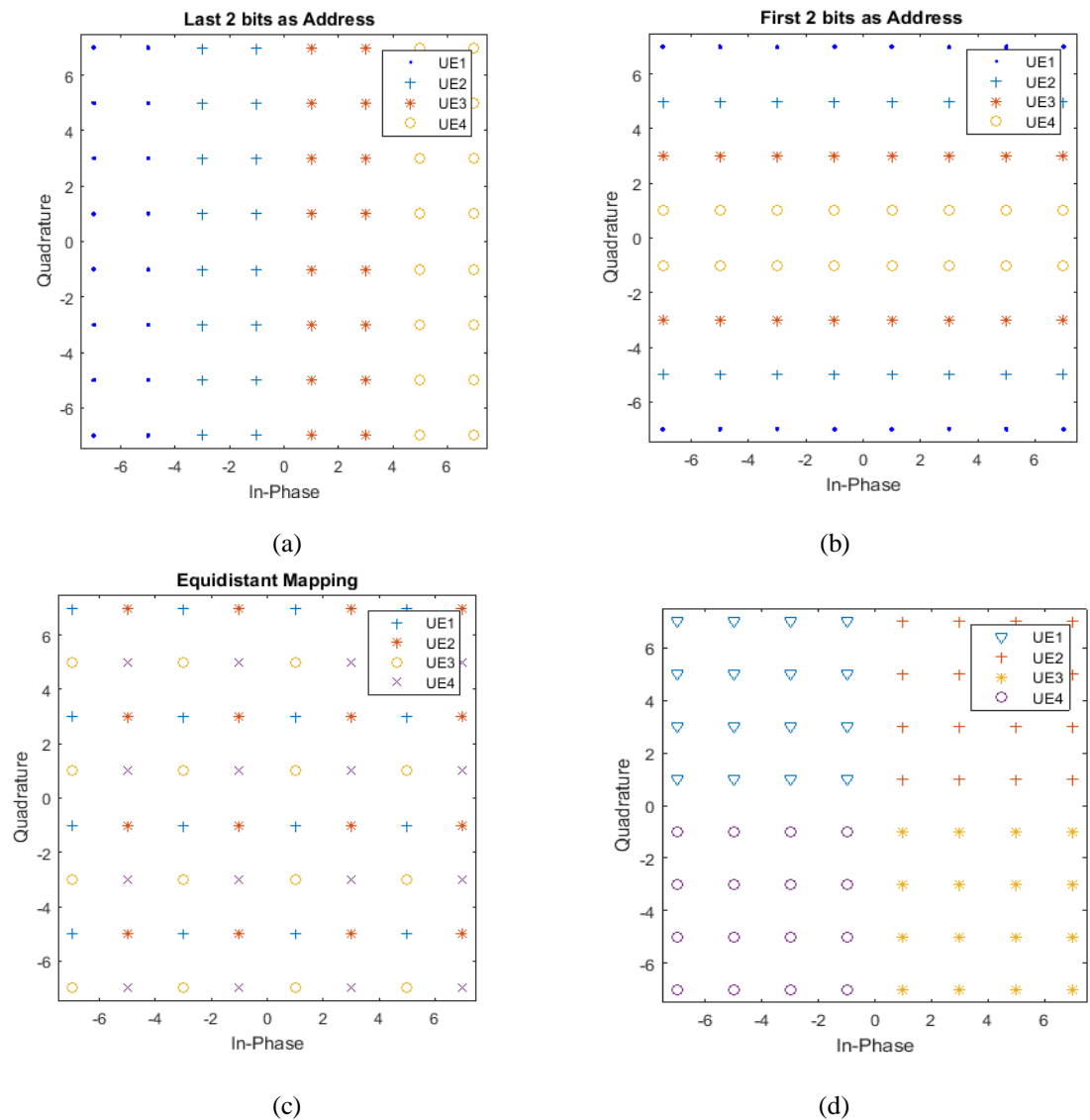

(a)     (b)

(c)     (d)

Figure 8. Several possible means of 64 QAM constellation sharing among 4 users

(a) Last 2 bits out of 6  are address bits     (b) First 2 bits out of 6  are address bits
(c) Equidistant mapping based on LUT     (d) First and 4th are address bits

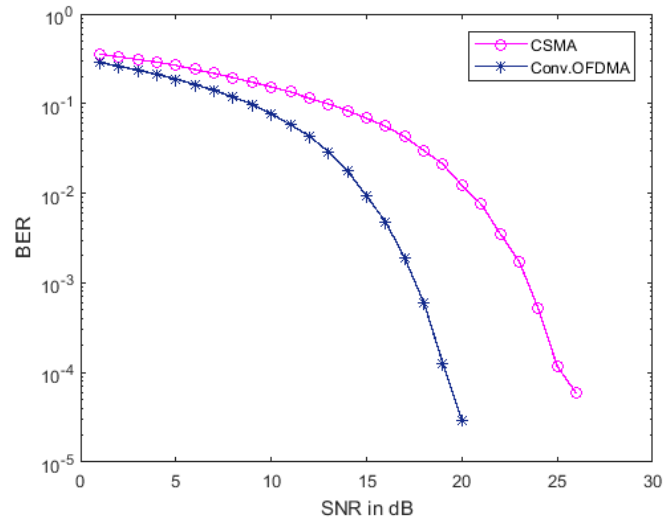

Figure 9. BER performance comparison of the proposed CSMA scheme with conventional OFDMA with single user allocation per RB

As explained in earlier section, it is seen that BER performance deteriorates gradually with increasing number of users that share the constellation space per resource block. Figure 10 shows the impact of accommodating more number of users in a constellation validating the equation 5. The reason for this is further revealed in the constellation plots given in figure 11 simulated for a normalized but relatively lower SNR of 20 dB. While the single user per RB has its constellation points much apart from each other, those with 4 users per RB are likely to suffer [31]. While the allocation of an RB to single user with 64 QAM modulation has resulted in good BER performance, a relative decay in BER performance is observed as the number of users sharing the RB increase as observed in a 4 and 8 users per RB respectively, resulting in higher SNR requirements.

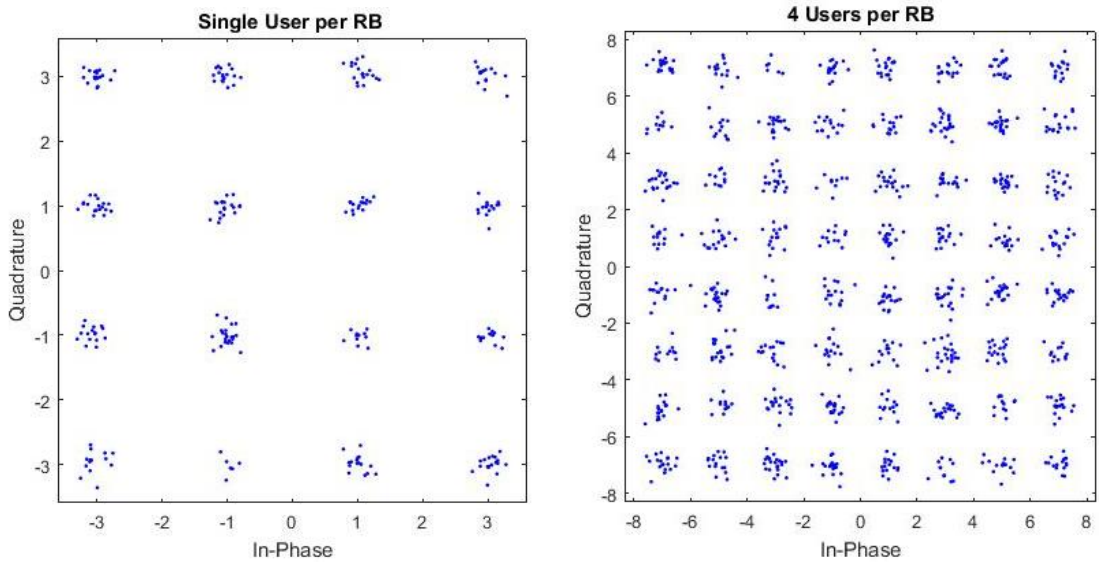

Figure 10. Constellation comparison of single user allocation per RB (left) with same RB shared by 4 users (left)

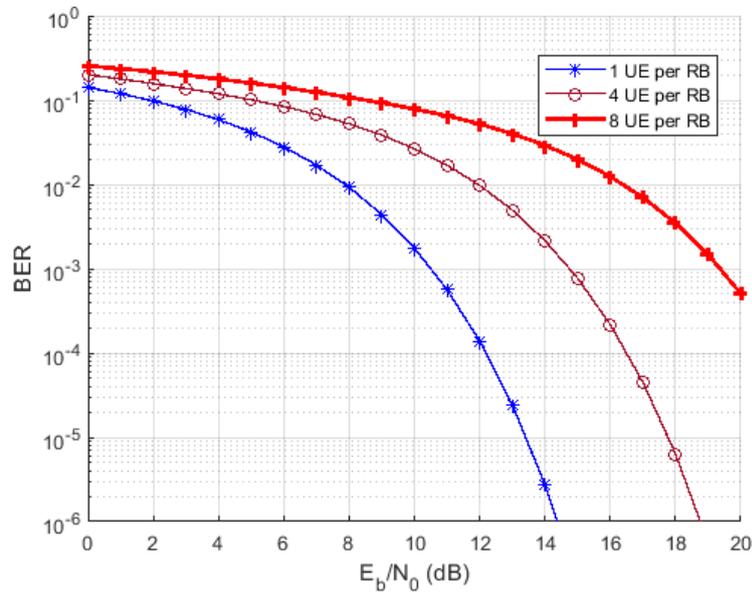

Figure 11. BER performance comparison of the scheme for single user, 4 user and 8 user sharing single RB employing 64 QAM OFDM

## 5. CONCLUSION

This paper presented a NOMA scheme to increase the simultaneous user capacity, especially for mMTC scenario. The scheme worked by sharing a single RB among multiple users by separating them in constellation domain, which was a slight modification to OFDMA based 5G NR approach where the entire block was allocated to single user only. The means of separating the users in constellation domain and the means of sharing the symbols in RB was depicted. Analysis on the increase in user capacity, its impact on throughput and the BER was conducted. Simulation results showed the various constellation allocation and performance comparison of the scheme with conventional transmission. The scheme is capable providing the much required increase in user capacity with slight modification in existing LTE or 5G NR structure. The simplicity with which the users share the same RB and the trivial means of differentiating their corresponding data after reception never necessitates the need for complex receivers. These features make the CSMA scheme a strong contender for the 5G mMTC. The possibilities to provide different QoS, optimum constellation mapping to achieve shaping gain and better PAPR need to be explored in future.

**CONFLICTS OF INTEREST**

The authors declare no conflict of interest.

## AUTHORS

**Kiran V. Shanbhag** received his BE degree in Electronics and Communication Engineering from Anjuman Institute of Technology and Management, Bhatkal, Karnataka, India in 2005 and received his MTech degree in Communication Engineering from National Institute of Technology Karnataka, Surathkal, India in 2010. He is currently pursuing PhD in Department of Electronics and Communication Engineering at St Joseph Engineering College, Mangaluru, India. His research interests include digital signal processing for communication, multicarrier modulation schemes and MIMO radar. He has 4 journal papers and 3 conference papers to his credit. Currently he is assistant professor at department of electronics and communication engineering, Anjuman Institute of Technology and Management, Bhatkal, Karnataka, India.

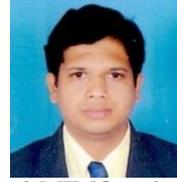

**Savitha H. M**. received her BE degree in Electronics and Communication Engineering from P.E.S. College of Engineering, Mandya, Karnataka, India in 1987 and received her MTech degree in Digital Electronics and Communication from NMAMIT, Nitte, Karnataka, India in 2001. She got her PhD in Communication Engineering from National Institute of Technology Karnataka, Surathkal, India in 2014. She was associated with St Joseph Engineering College, Mangaluru as Professor and has a teaching experience of 22 years. Her research interests include error control coding, OFDM, wireless communication and signal processing. She has 7 international journals and 13 international/national conferences to her credit.

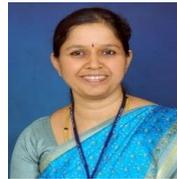